\documentstyle[prl,aps]{revtex}

\begin{document}
\columnsep0.1truecm
\draft
\title{Elastic Theory Has Zero Radius of Convergence}
\author{Alex Buchel and James P. Sethna}
\address{Laboratory of Atomic and Solid State Physics,\\
Cornell University, Ithaca, NY, 14853-2501}
\maketitle

\begin{abstract}

Nonlinear elastic theory studies the elastic constants of a material
(such as Young's modulus or bulk modulus) as a power series in the
applied load.  The inverse bulk modulus K, for example depends on 
the compression  P: 
$ {1/ K(P)} = c_0 + c_1 P + c_2 P^2 \cdots + c_n P^n + \cdots $.
Elastic materials that allow cracks are unstable at finite temperature with 
respect to fracture under a stretching load; as a result,
the above power series has zero radius of convergence and thus
can at best be an asymptotic series.
Considering thermal nucleation of cracks in a two-dimensional 
isotropic, linear--elastic  material at finite temperature we compute  
the asymptotic form  $ { c_{n+1}/ c_n}\to C n^{1/2}$ as $n \rightarrow \infty$.
We present an explicit formula for $C$ as a function of temperature
and material properties.

\end{abstract}


\widetext

Hooke's law, $F=k x$, representing the elastic response of a body to an
external stress, is only the first term in a Taylor's series.  This
paper addresses the convergence properties of this expansion.  We argue
generally that the radius of convergence of elastic theory is zero, and
for two-dimensional isotropic linear-elastic theory allowing for brittle 
fracture, we can calculate the asymptotic behavior of the coefficients.

For simplicity, let's consider the bulk modulus $K(P)$, 
\begin{equation}
{1 \over K(P)} = -{1 \over V} \left({\partial V \over \partial P}\right)_T 
		= c_0 + c_1 P + c_2 P^2 \cdots + c_n P^n + \cdots
\end{equation}   
Under stretching $(P<0)$, the true ground state is fractured into
pieces (relieving the strain energy).  As a result, $P=0$ can not be 
a point of analyticity for $K(P)$, and thus (1) has zero radius of 
convergence.

Similar arguments were used by Dyson\cite{dyson} in 1952, where
he argued that calculations in quantum electrodynamics, expressed 
as a power series in the fine structure constant $e^2/\hbar c \approx 1/137$, 
had zero radius of convergence (because negative values of $e^2$ lead to 
unstable theories).  This did not prevent these calculations
from being useful (indeed, they represent the best quantitative agreement
between theory and experiment known to science). The community  believes  
these expansions are asymptotic in the same sense as Stirling's approximation
 $\Gamma(n) =(n-1)!\sim\rm{e}^{-n}n^n{\biggl({{2\pi}/
 n}\biggr)}^{1/2}\biggl({1+{1/(12 n)}+{1/(288n^2)}+\cdots}\biggr)$:
at any fixed $n$ no matter how large, Stirling's
series in $1/n$ eventually diverges, but the difference between the
function and the $M^{th}$ approximation goes to zero faster than $1/n^M$
as $n \to \infty$.  

Since Dyson's work, field theoretic methods have been 
developed\cite{{f1},{f2},{f3},{f4}} to relate the instabilities in the
theories at small negative couplings to the high--order terms in
perturbation theory.  Here we apply these methods to a particular
case, using the statistical mechanics of thermally nucleated cracks 
to calculate the high--order terms $c_n$ in the inverse bulk modulus (1).
Statistical and thermodynamic approaches to crack nucleation and
fracture have an established history\cite{selinger}.
However, most work in this area is concentrated on failure at rather
high stresses, near the threshold for instability (the spinodal point).
The high--order terms in the perturbative expansion of the 
inverse bulk modulus are governed by the elastic response of the
material to infinitesimally small tension (see below), so we are far 
away from the spinodal point and linear elastic theory is an adequate 
description.          

Consider an infinite two-dimensional isotropic linear-elastic material
subject to uniform compression $P$ at infinity.  Creation of a
cut of length $\ell$ will increase the energy by $2 \alpha \ell$, where
$\alpha$ is the surface tension (the energy per unit length of edge),
with a factor of $2$ because of the two free surfaces. On the other
hand, for negative $P$ (uniform tension) the cut will open up because
of elastic relaxation.  Calculating this relaxation energy we find the 
total energy $E$ of a crack of length $\ell$:
\begin{equation} E(\ell)=2 \alpha \ell -{{\pi P^2 (1- \sigma^2) \ell^2} \over
{4 Y}}.
\end{equation}
Introducing 
\begin{equation}
\ell_c={{4 Y \alpha} \over {\pi P^2 (1- \sigma^2)}}
\end{equation}
we can rewrite the energy of the crack as
\begin{equation}
E=2 \alpha \ell -\alpha {\ell^2 \over \ell_c}.
\end{equation}

It follows that cracks with $\ell>\ell_c$ will grow, giving rise to
the fracture of the material, while those with $\ell<\ell_c$ will heal
--- a result first obtained by Griffith\cite{c1}.
This is the instability that is responsible for the breakdown of elastic
perturbation theory.  Because the energy $E(\ell_c) = \alpha \ell_c$
grows as $1/ P^2$ as $P \rightarrow 0$, interactions between 
thermally nucleated cracks are unimportant at small $P$ and low temperatures
(allowing us to use the ``dilute gas approximation'').

The thermodynamic properties of a macroscopic system can be obtained 
from its partition function $Z$:
\begin{equation}
Z=\sum_n \exp (-\beta E_n)
\end{equation}     
where the sum is over all states of the system. Once a perturbative expansion 
for the free energy $F =-{(1 /  \beta)} \ln Z$ is known, one can calculate
the power series expansion for the inverse bulk modulus using  
\begin{equation}
{1 \over K(P) }=-{1 \over {P A }}{\biggl ({{\partial F} \over
{\partial P}}\biggr)_T}
\end{equation}
where the elastic material has area $A$.

For $P<0$, our model is in a metastable state, and direct computation of
the partition function should yield a divergent result. A similar
problem for the three-dimensional Ising model was solved 
by Langer\cite{langer}: one has
to compute the partition function in a stable state $P>0$, and then do
an analytical continuation in parameter space to the state of interest.
The free energy develops an imaginary part in the unstable state,
related to the decay rate for fracture\cite{langer2}: the situation is
similar to that of barrier tunneling in quantum mechanics \cite{affleck}, 
where the imaginary part in the energy gives the decay rate of a resonance.

The calculation of the imaginary part of the partition function is
dominated by a saddle point, that in our case is a straight cut of 
length $\ell_c$.  The straight cut is the saddle point because 
it gains the most elastic relaxation energy for a given number of broken
bonds (as can be checked with a direct calculation\cite{we}).

For simplicity, we start by considering the model without including
the quadratic fluctuations around the saddle point.  The partition
function for a dilute gas of straight cuts of arbitrary length at their
equilibrium shape for the tension $P$ can be calculated directly: the 
imaginary part of the free energy for $P<0$ is
\begin{equation}
{\rm Im} F^{simple}(P)={2 \over {\beta  |P|} }{\biggl( {Y \over {\beta 
 \lambda^2 (1- \sigma^2)}} \biggr)}^{1/2}
\biggl ({2 \pi {{A} \over {\lambda ^2}}} \biggr )\exp{ \biggl\lbrace  
{{-4 \beta Y
 \alpha^2 } \over { \pi P^2 (1- \sigma^2)}  } \biggr\rbrace}
\end{equation}
with $\lambda$ being the ultraviolet cutoff in the theory (roughly, the 
interatomic distance).  (The factor $(2 \pi A/{\lambda^2})$ comes from
the zero modes for rotating and translating the cracks.)  

Assuming the free energy is analytic in the complex $P$ plane except for
a branch cut along $P \in (-\infty , 0]$,
we obtain a Cauchy representation \cite{{f2},{f4}} for the free energy
\begin{equation}
F(P)={1 \over \pi}\int\limits_{-\infty}^0{{{{\rm Im} F(P')} \over {P'-P}} dP'}.
\end{equation}
As was first established for similar problems in field 
theory\cite{{f1},{f2},{f3}}, this relation determines the high--order 
terms in perturbative expansion of the free energy $F(P)=\sum_n {f_n P^n}$
\begin{equation}
f_n={1 \over \pi}\int\limits_{-\infty}^0{{{\rm Im} F(P')} \over
{{P'}^{n+1}}}dP'
\end{equation} 

Because the saddle point calculation becomes more and more accurate
as $P \to 0$, and because the integrals in equation (9) are dominated
by small $P$ as $n \to \infty$, using the saddle--point form for the
imaginary part of the free energy yields the correct $n\to\infty$ asymptotic 
behavior of the high-order coefficients $f_n$ in the free energy. For
our initial approximation~(7)
\begin{equation}
f_n^{simple}={(-1)}^{n+1}\Gamma{\biggl({{n+1}\over
2}\biggr)}{\biggl({{\pi(1-\sigma^2)}\over{4\beta Y
\alpha^2}}\biggr)}^{n/2}\biggl ({2 \pi {{A} \over {\lambda ^2}}}
\biggr ){1\over{2\pi^{1/2}\beta^2\alpha\lambda}}
\end{equation}
We can then use the thermodynamic relation (6) to show 
${c_{n}} = -(n+2)f_{n+2}/A$, 
and thus calculate the asymptotic behavior of the expansion of the
bulk modulus:
\begin{equation}
{c_{n+1} \over c_n}\rightarrow 
	- n^{1/2} {\biggl({{{\pi (1-\sigma^2)} \over {8 \beta Y
\alpha^2}}\biggr)}^{1/2}} \mbox{\hspace{0.1in}  as n} \rightarrow\infty
\end{equation}
which indicates that the high--order terms $c_n$ in the perturbation 
expansion for the inverse bulk modulus roughly grow as $(n/2)!$.  
The formula (11) will still be correct once we add the quadratic fluctuations 
(although $\alpha$ will develop temperature--dependent corrections).
For cracks in three dimensions, we expect the scaling 
${c_{n+1} / c_n}\sim n^{1/4}$ as $n \to \infty$, using similar arguments.

The above calculation ignores the quadratic fluctuations about the
saddle point, which will change the prefactor in the expression (7) for
the imaginary part of the free energy and will renormalize the surface
tension $\alpha$.There are two kinds of quadratic fluctuations we have
to deal with. (I) {\it Curvy cuts} --- changes in the shape of the tear
in the material: deviations of the broken bonds from a straight-line
configuration. (II) {\it Surface phonons} --- thermal fluctuations of
the free surface of the crack about its equilibrium opening shape. Since
the energy of the curvy cracks is calculated for their equilibrium
configuration, the response of the surface phonons to curving the cut
is already incorporated, so the quadratic fluctuations (II) can be
calculated independently from (I).

In both cases the theory needs regularization: we must decide exactly
how to introduce the ultraviolet cut--off $\lambda$.  We describe the
curvy crack as a bunch of line segments, parameterized by kink angles
$\alpha_i$ --- angles between consecutive segments.  We calculate the energy
release in the material with a crack using a conformal mapping from the
plane with a cut to the exterior of a unit circle \cite{m}, which for a
piecewise linear cut is given by a modified Schwartz-Christoffel map.
It is convenient for the regularization of the surface phonons to use
equally spaced points on the unit circle, of angular spacing 
$\pi \lambda / \ell$; this corresponds to points which accumulate
at the edges of the original crack cut.  For consistency, we use the 
same regularization for the curvy crack segments.  We have also used
the (less convenient, more natural) regularization of equally spaced points 
along the original crack cut, and the results are similar (although
the convergence is not as convincing): the form of the
imaginary part of the free energy and the ratio of the
high--order elastic coefficients (11) do not change, but (see below)
the constants $s_0$, $s_1$ and $s_2$ and the temperature--dependence 
of $\alpha$
do change \cite{we}.

To calculate the quadratic fluctuations due to the surface phonons, 
we must calculate the determinant of the matrix $M_p$, given 
by the energy spectrum of the normal modes about the crack's 
equlibrium (saddle point) opening: 
\begin{equation}
det M_p = 2 \pi {\biggl({{2 (1-\sigma^2)}\over{\beta Y \lambda^2}
}\biggr)}^2 n \exp{\biggl\lbrace{n\biggl({ 2\ln{{\beta Y \lambda^2}
\over{2 (1-\sigma^2)}}-2}\biggr)}\biggr\rbrace}
\end{equation}
where $n$ is number of kinks. Similarly, the quadratic fluctuations 
due to curving the crack is given by  the determinant of the 
matrix $M_{ij}$ that specifies the change in energy release to
quadratic order in the kink angles $\alpha_i$ (so the
energy change due to the curving of a straight cut $\Delta
E(\{\alpha_i\}) = M_{ij} \alpha_i \alpha_j + O(\alpha^3)$). We have
obtained an analytical expression for the entries of the above matrix by
calculating, to quadratic order, the energy of a crack with two kinks.
This lengthly and tedious calculation \cite{we} was checked both using a
finite--element crack simulation \cite{franc2d} and an exact
analytical solution for a cut that is an arc of a circle. We assume that  
scaling of the curvy crack determinant is of the same form 
as for the surface phonons : 
\begin{equation}
det M \sim  s_0
n^{s_1}\exp{\biggl\lbrace{n\biggl({\ln{{\beta\alpha\lambda
}\over{4 \pi^3}} + s_2}\biggr)}\biggr\rbrace}
\end{equation}
as the number of kinks $n \to \infty$. We extract the constants $s_0$, $s_1$, 
and $s_2$ numerically, using
systems with up to $n=400$ kinks: we extrapolate visually to find 
$\ln s_0=1000\pm1000$, $s_1=-400\pm100$, and $s_2=7.5\pm0.5$. Notice that 
$s_0$ is subdominant to $s_1$, which is subdominant to $s_2$, so our errors 
get rather large - even without including our uncertainty as to the scaling 
form (13). 

These quadratic fluctuations modify the imaginary part of the
free energy (7) as follows:
\begin{eqnarray}
{{\rm Im} F(P)} ={s_0\over\sqrt{8\pi}} 
{{\beta Y \lambda^2}\over{1-\sigma^2}}
{\biggl({{l_c \over \lambda}}\biggr)}^{-(1+s_1)/2}\mbox{\hspace{2.0in}}  \\
\mbox{\hspace{1.0in}}\exp\biggl\lbrace
{ {l_c \over \lambda}{\biggl( {1 - {s_2 \over2} + \ln
{{2(1-\sigma^2)}\over{\beta Y \lambda^2}} + {1\over2}\ln{{4\pi^3}\over
{\beta\alpha\lambda}}}\biggr)}}\biggr\rbrace {{\rm Im} F^{simple}(P)} 
\nonumber
\end{eqnarray}
with $l_c$ given by (3) and ${\rm Im}F^{simple}(P)$ given by (7). 
The first term in (14) modifies the prefactor, 
the other one effectively renormalizes the surface tension: 
\begin{equation}
\alpha_r = \alpha +{1 \over {2\beta \lambda}}\biggl( {{s_2\over2} - 1 +
\ln{{\beta Y \lambda^2} \over {2 (1-\sigma^2)}} + {1\over2}\ln
{{\beta\alpha\lambda}\over{4\pi^3}}}\biggr)  
\end{equation}

Finally, we derive the correct asymptotic form for $c_n$:
\begin{equation}
{c_n}=(-1)^n (n+2) \Gamma{\biggl({{n-s_1}\over2}\biggr)}
{\biggl({{\pi(1-\sigma^2)}\over{4\beta Y
\alpha_r^2}}\biggr)}^{n/2}\biggl ( {{{2 \pi} \over {\lambda ^2}}}
\biggr ){\biggl( {{\beta \alpha_r^2
\lambda}\over\alpha}\biggr)}^{s_1/2}{{s_0 Y \lambda} \over
{8\sqrt{2\pi}\beta\alpha (1- \sigma^2) }}
\end{equation}
This is the main result of our paper.

We warn the reader to treat our results in the proper context. First, we do
not expect this calculation to have experimental implications in the
near future.  Real fracture invariably occurs on inhomogeneities in the
material: pre-existing surface or bulk microcracks, dislocation tangles,
or grain boundaries.  Even for a perfect dislocation-free crystal with
stabilized surfaces, the effects we describe will remain immensely
small: the reason that incredibly tiny compression make (relatively)
large changes in volume (leading to large high-order terms in the power
series) is because they suppress the (already incredibly rare) opening
of large thermally nucleated cracks: measuring these effects would seem
infeasible.

Second, our results can be viewed as a straightforward extension to the
solid-gas sublimation point of Langer \cite{{langer},{langer2}} and
Fisher's \cite{Fisher} theory of the essential singularities at the
liquid-gas transition.  Indeed, if we allow for vapor pressure in our
model, then our system will be in the gas phase at $P=0$.  The essential
singularity we calculate shifts from $P=0$ to the vapor pressure.
If we measure the nonlinear bulk modulus as an expansion about (say)
atmospheric pressure, it could converge: but the radius of convergence
would be bounded by the difference between the point of expansion and the
vapor pressure.

Third, we have forbidden dislocation nucleation and plastic flow in our
model.  Dislocation emission is crucial for ductile fracture, but 
by restricting ourselves to a brittle fracture of defect--free
materials we have escaped many (lethal) complications.  Dislocations
are in principle important: the nucleation\cite{nelson}
barrier $E_{dis}$ for two edge dislocations in an isotropic linear--elastic
material under uniform tension $P$ with equal and opposite Burger's
vectors $\vec b$ is
\begin{equation}
E_{dis}={{Y b^2}\over {4 \pi (1-\sigma^2)} }\ln {Y\over P} + E_0 
\end{equation}
where $E_0$ is a $P$ independent part that includes the dislocation 
core energy.
The fact that $E_{dis}$ grows like $1/\ln P$ as $P\to 0$ (much more
slowly than the corresponding barrier for cracks) tells that 
in more realistic models dislocations and the resulting plastic 
flow\cite{ambegaokar} cannot be  ignored.
While dislocations may not themselves lead to a catastrophic instability
in the theory (and thus to an imaginary part in the free energy?),
they will strongly affect the dynamics of crack nucleation as noted
above.

Fourth, we are being careful to distinguish tension from negative
hydrostatic pressure.  The response of a crystal to a force depends in a
fundamental way on whether the force couples to the atoms or to the
lattice (i.e., to the broken symmetry).  Under gravity, all crystals
will flow like anisotropic liquids, with a viscosity given simply in
terms of the vacancy diffusion constant.  In contrast, a crystal gripped
at the sides and strained will not flow with a rate linear in the
external force: under a strain produced by coupling to the surface
layers, single vacancies cannot individually relieve the strain, and a
dislocation pair (a loop in a three-dimensional case) must appear in the
material for strain relief to occur: the plastic response of a
two--dimensional crystal to external strain is a temperature--dependent
power--law in the external force\cite{ambegaokar}.  Although we call the
compression (or tension) $P$, it is not hydrostatic pressure, but a
coupling to the lattice.  Under negative hydrostatic pressure, vacancies
will have a negative chemical potential $\mu$, and the dominant fracture
mechanism becomes the nucleation of vacancy clusters or voids (rather
than Griffith-type critical microcracks), as noted by Golubovi\'c and
collaborators\cite{golubovic}. For negative chemical potential $\mu$ per
unit area, their vacancy clusters (up to constants) have energy
$E_{vac}(R) \sim \alpha R - |\mu| R^2$. If we identify $\mu$ with $P$,
comparing with (2) we see that the vacancy cluster gains an energy
linear in $P$, while the crack gain is quadratic.  This leads to a
problem that maps onto Langer's calculation, leading to the asymptotic
relation $c_{n+1}/c_{n}\sim n$, so the coefficients in this case would
diverge more strongly: $c_n \sim n!$. However, the identification of
$\mu$ with $P$ demands a mechanism for relieving elastic tension by the
creation of vacancies.  Ignoring possible effects of the boundaries
(presumed infinitely far removed), as noted above vacancies will be
created (and hence relieve tension) only through dislocation motion,
which we have explicitly excluded from our model.

We acknowledge the support of DOE Grant DE-FG02-88-ER45364. 
We would like to thank Yakov Kanter, Eugene Kolomeisky, 
Paul Houle, Tony Ingraffea,
Paul Wawrzynek, and Robb Thompson for useful conversations.

\end{document}